\documentclass[aps,pre,twocolumn,amsmath,amssymb,amsfonts]{revtex4}
\usepackage{bbold}
\usepackage{epsfig}
\usepackage{graphicx}
\usepackage{dcolumn}
\usepackage{bm}
\usepackage{xcolor}
\usepackage{bm}

\usepackage[titletoc,title]{appendix}
\begin{document}
\author{Itzhak Fouxon$^{1,2}$}\email{itzhak8@gmail.com}
\author{Yizhar Or$^{2}$}\email{izi@technion.ac.il}
\affiliation{$^1$ Department of Chemical Engineering, Technion, Haifa 32000, Israel}
\affiliation{$^2$ Faculty of Mechanical Engineering, Technion - Israel Institute of Technology, Haifa, 3200003, Israel}
\title{Inertial self-propulsion of spherical microswimmers by rotation-translation coupling}

\begin{abstract}

We study swimming of small spherical particles who regulate fluid flow on their surface by applying tangential squirming strokes. We derive translational and rotational velocities for any given stroke which is not restricted by
axial symmetry as assumed usually. The formulation includes inertia of both the fluid and the swimmer, motivated by inertia's relevance for large Volvox colonies. We show that inertial contribution to mean speed comes from dynamic coupling between translation and rotation, which occurs only for strokes that break axial symmetry. Remarkably, this effect enables overcoming the scallop
theorem on impossibility of propulsion by time-reversible stroke. We study examples of tangential strokes of axisymmetric travelling wave, and of
asymmetric time-reversible flapping. In the latter case, we find that inertia-driven mean speed is optimized for flapping frequency and swimmer's size which fall well within the range of realistic physical values for Volvox
colonies. We conjecture that similarly to Paramecium, large Volvox could use time-reversible strokes for inertia-driven swimming coupled with their rotations.

\end{abstract}

\maketitle

\section{Introduction}

Spherical microswimmers is a unique model of swimming at low Reynolds number which is theoretically tractable.  Its introduction is motivated by ciliated microorganisms \cite{Lighthill1952,Blake} as well as flagellated colonies of
Volvox algae \cite{Goldstein,review}. The interaction of these organisms with the fluid is described by no-slip boundary condition on nearly spherical time-dependent envelope of the tips of cilia or flagella whose motion is actuated by the swimmer
(squirming). 
This boundary velocity sets the fluid in motion that applies propulsion force on the swimmer \cite{Goldstein,review,Lighthill1952,Blake,Brumley1,ishimoto,ishyam,Brumley2,Brumley3,Brumley4,swirl,fhos,paklauga,Felderhof,Felderhofco,gonzlauga,ardekani}. Thanks to smallness of the Reynolds number, the fluid motion can be described by linear, steady or unsteady, Stokes equations. The linearity and the spherical shape open the opportunity of detailed solution connecting microscopic motions of the actuated envelope with the swimmer's motion as a whole.

Previous treatments mostly neglected inertias of both the fluid and the swimmer and assumed axially symmetric swimming stroke as in the original formulation  \cite{Lighthill1952,Blake}. This leads to 
substantial simplifications in the coupled system of the Navier-Stokes equations (NSE) governing the flow, and the Newton equations governing the swimmer’s motion.
We first consider the neglect of fluid inertia using which the convective and time derivative terms of the NSE are dropped. The convective term in the equations, giving flow derivative along the streamline, is
negligible since the Reynolds number (defined as the product of the swimmer's size and velocity divided by the kinematic viscosity) is less than $0.1$ even for largest Volvox colonies \cite{swirl}. In contrast, the neglect of time-derivative term in the NSE requires careful consideration. If the fluid inertia is negligible then this term can be discarded and steady Stokes equations of the flow apply \cite{Lighthill1952,Blake,purcell,sw}. However, this neglect is invalid for largest Volvox colonies of hundreds of microns in size (Volvox comes in aggregates of different numbers of cells and has a range of sizes) \cite{Goldstein,swirl}. Indeed, significance of inertia of the fluid is determined by the Roshko number $Ro=\sigma\tau_d$ where $\sigma$ is the frequency of the periodic stroke and $\tau_d$ is
the characteristic time of outward viscous diffusion of momentum from the sphere (sometimes $Ro$ is called oscillatory Reynolds number \cite{ishimoto,lauga0}). This time is proportional to the squared radius of the swimmer
divided by the kinematic viscosity. Using the experimentally observed value of $\sigma=203$ radian per second \cite{Goldstein} we have $Ro\sim 1$ for a colony with radius of hundred microns. This would invalidate the neglect of
inertia for Volvox since the typical size range is between one and five hundred microns. In fact, we demonstrate below that it is plausible that the definition of $Ro$ must be multiplied by a numerical factor of $1/9$. Thus it
is reasonable that for colonies smaller than $300$ microns we can assume $Ro\ll 1$ (we observe that $Ro$ has fast quadratic dependence on the radius). For these small colonies the momentum redistributes over the fluid
before the swimmer moves the flagella significantly. The flow is then steady Stokes flow determined by instantaneous position and velocity of the flagella. However, for colonies larger than $300$ microns where $Ro>1$,
momentum diffusion and swimming stroke are coupled non-trivially. This coupling is described by the unsteady time-derivative term in the NSE whose inclusion is necessary. This term brings memory where the propulsion force is
determined not only by instantaneous stroke but also by its past values \cite{maxeyriley,Basset,boussi,ll6}. Though this sometimes does not influence the time-averaged net propulsion velocity for periodic strokes, it is relevant
for nutrient uptake \cite{ardekani} and reaction to external stimuli \cite{para}. It must be stressed here that the $300$ microns threshold is only an estimate. Detailed comparison between the theory and the experiment
is needed in order to decide which colonies can be considered inertialess and which must be treated as inertial.

The second of the usually made assumptions is the neglect of swimmer's inertia 
implying that the motion occurs at zero force and torque. This is also invalid for large colonies where time-scale of viscous drag is comparable with the period of the stroke. In fact, densities of the swimmer and of
the fluid are very close so the relevance of fluid and particle inertias is determined by the same parameter $Ro$. Finally, the third assumption of axially symmetric stroke, in the traditional form without the swirl \cite{swirl},
allows to use the general solution for axially symmetric flow. This assumption brings zero rotation thus removing the characteristic rotation of Volvox that gave it its name. The rotational velocity was included
without destroying the symmetry by introducing axisymmetric swirl component \cite{swirl}. In this framework there is no coupling between translation and rotation (in sharp contrast with the coupled rotation and translation
considered here). Thus, even though introduction of the axisymmetric swirl helps fitting the rotational velocity, it does not help fitting the translational velocity that is independent of parameters of the swirl. Consequently, the
existing discrepancy between theoretical predictions for the swimming velocity and the data cannot be resolved by the axisymmetric swirl \cite{swirl}. Axially symmetric stroke also restricts the motion to a straight line
which is not always the case \cite{paklauga}. This stroke can be unsuitable for description of physical phenomenona as in the case of phototaxis where a complex interaction between rotation and translation regulates turning of the colony towards light source \cite{photo,paklauga}.

The three assumptions considered above can be invalid, either because $Ro\sim 1$ or because of asymmetry of the strokes. Few previous works relaxed some of the assumptions. The work \cite{gonzlauga} studied the limit case of inertial
heavy (called dense in \cite{gonzlauga}) swimmer in inertialess fluid, disregarding rotation of tangential squirmer. In \cite{ardekani} the equation of motion for translational velocity of inertial tangential squirmer in inertial
flow was developed using the unsteady reciprocal theorem. However, the equation was solved only for axially symmetric case where both inertias do not change the
time-averaged velocity, and the motion is pure axial translation without rotation. This results in oscillatory terms that give nonzero transient motion for a maneuver starting from rest, but no net propulsion in steady state \cite{ishimoto}.
On the other hand, stroke's symmetry assumption was relaxed in \cite{fhos,paklauga,Felderhof,Felderhofco} where inertial effects were not considered.

In this work, we relax all the three assumptions above, which are negligible inertias of the fluid and the swimmer and the axisymmetric swimming strokes. These assumptions do not necessarily hold in practice. The Roshko number
is not small for large colonies where the net force is nonzero and the stroke can depend on the azimuthal angle as in phototaxis
\cite{photo,paklauga}. In contrast, we continue using the assumption of small Reynolds number which is satisfied in practice (finite Reynolds number corrections become relevant for sizes close to one millimeter
\cite{khair,spelauga}). For making the arising problem tractable we make a simplifying assumption that the swimming stroke is tangential. This assumption does not hold for Volvox, however it could be
that the discovered qualitative phenomenon has relevance for motion of the algae. Namely, we demonstrate that interplay of the fluid and swimmer's inertia with tangential asymmetric stroke gives a previously unstudied mechanism of swimming.
In particular, it is usually believed that tangential squirmers that keep constant spherical shape cannot swim by a time-reversible stroke - the time-averaged velocity is zero \cite{ardekani,gonzlauga,ishimoto}.
However, this conclusion was made only for axially symmetric stroke that involves no rotation. We demonstrate that asymmetric time-reversible tangential stroke that does involve rotation can generate self-propulsion via nonlinear rotation-translation coupling. The net swimming velocity considered as a function of inertia parameter $Ro$ reduces to zero as $Ro \to 0$, in accord with the scallop theorem \cite{purcell}. As we show below, the leading order term at $Ro\ll 1$ is proportional
to time-averaged vector product of instantaneous translational and rotational velocities at $Ro=0$. This result is independent of the details of the stroke. We present a concrete example of a time-reversible stroke with nonzero
average nondimensional propulsion speed at any $Ro>0$. The nondimensional 
speed decays for small and large values of $Ro$, and reaches a maximum at an intermediate value. This value falls within feasible physical values of large Volvox colonies. Thus the discovered phenomenon might have relevance
for these algae.
For small swimmer's size with negligible inertia, our formulas for translational and rotational velocity of inertial spherical tangential squirmer in inertial fluid in terms of arbitrary swimming stroke reduce to those of
\cite{stonesamuel} for inertialess squirmer in inertialess fluid.

We demonstrate that inertia changes the swimming velocity, obtained by neglecting inertia, directly, by producing additive correction, and indirectly by changing the rotation velocity. The direct contribution has zero time-average. Thus scallop theorem implies that swimmer who does not rotate or whose rotation axis is aligned with the translational velocity could not swim by time-reversible stroke as was found in \cite{ardekani,gonzlauga,ishimoto} by direct calculation. It must be stressed though that translation during one period in these cases can be non-zero. The indirect influence through the rotation can change the net velocity through the non-linear interplay of the stroke and the spatial rotation (rotation-translation coupling). This makes the motion quite different from the inertialess case and enables overcoming the scallop theorem.

In the next Section we demonstrate that insight in the motion can be obtained by separating the flow into the inertialess component due to the swimming stroke and inertial friction component. The former describes the motion of inertialess swimmer in inertial fluid and causes the corresponding translational and swimming velocities. The frictional component describes that, due to inertia, the swimmer has finite (frequency dependent) relaxation times and lags behind the velocities of inertialess component.

\section{Flow as sum of inertialess swimming and inertial components}

In this Section we introduce decomposition of the flow. One component of the flow describes the result of force- and torque-free swimming of inertialess swimmer in inertial fluid. This flow is of separate interest. The other component describes the propulsion force (and torque) that acts to produce frequency-dependent relaxation of the swimmer's velocity to the velocity of inertialess swimmer.

We consider tangential squirmers that have constant shape of sphere with radius $a$. The swimming stroke produces in the body-fixed coordinate frame a purely tangential motion of the spherical surface at
velocity $\bm v_b(\bm x, t)$. Thus at time $t$ velocity of the material point whose coordinate in the body-fixed frame is $\bm x$ is $\bm v_b(\bm x, t)$ where the domain of definition of $\bm v_b$ is $|\bm x|=a$ and $\bm x\cdot\bm v_b(\bm x)=0$. Here in parallel with assumptions of Volvox modeling \cite{Goldstein} we assume that the fluid does not change the "flagella" motion and $\bm v_b(\bm x, t)$ is a given quantity determined
by the "biology of the organism". We stress that $\bm v_b(\bm x, t)$ is Eulerian and not Lagrangian flow field. In the often used (cf. \cite{Lighthill1952,swirl}) Lagrangian description one would describe the swimming stroke by running coordinates of material points $\theta(t)=\theta_0+f_1(\theta_0, \phi_0, t)$ and $\phi(t)=\phi_0+f_2(\theta_0, \phi_0, t)$
where $f_i(\theta_0, \phi_0, t)$ are given (small) functions characterizing the stroke as displacement fields of material points. Here $\theta_0$, $\phi_0$ are initial positions of the point on the sphere as in the usual
Lagrangian description of motion of the fluid \cite{ll6}. The velocity $\bm v_b(\bm x, t)$ is implicitly given by,
\begin{eqnarray}&&\!\!\!\!\!\!\!\!\!\!
\bm v_b(\theta_0\!+\!f_1, \phi_0\!+\!f_2, t)\!=\!a\hat {\bm \theta} \partial_t f_1\!+\!a\sin(\theta_0+ f_1)\hat {\bm \phi} \partial_t f_2,\label{impl}
\end{eqnarray}
where $\hat {\bm \theta}$, $\hat {\bm \phi}$ are unit vectors in the directions of growth of polar and azimuthal angles, respectively. Solving the equation for all $\theta_0$ and $\phi_0$ one can find $\bm v_b(\bm x, t)$ in
terms of $f_i$ (in practice this involves expansion in smallness of $f_i$, see e. g. \cite{stonesamuel}). Since both $\bm v_b$ and $f_i$ are determined by the stroke kinematics and not by the interaction with the fluid then we consider them
as equivalent descriptions which are assumed as given. Thus we will provide swimming velocities in terms of $\bm v_b(\bm x, t)$ that can then be rewritten in terms of $f_i$ by solving Eq.~(\ref{impl}).

We now consider description of the stroke in the rest frame of the fluid. The fluid frame coordinate of a material point with body-fixed coordinate $\bm x_b$ can be written as $\bm x_0(t)+\bm R(t)\bm x_b$. Here $\bm x_0(t)$ is the
position of the swimmer’s center in the frame of the fluid and the rotation matrix $\bm R(t)$ is orthogonal (the axes of the two frames can be made to coincide by translation of the origin and rotation). For definiteness we assume that at $t=0$
the frames coincide so that $\bm x_0(0)=0$ and $R_{ij}(0)=\delta_{ij}$. The equation on the time derivative of orthogonal matrix $\bm R( t)$ has the general form $\dot {R}_{ij}=\epsilon_{ikl}\Omega_kR_{lj}$ that defines the angular
velocity vector $\bm \Omega$. Using $\bm \Omega$ and center velocity $\bm v=d\bm x_0/dt$
we can write the fluid frame velocity of the material point with coordinate $\bm x_0(t)+\bm x$ where $|\bm x|=a$. This is given by the time derivative of $\bm x_0(t)+\bm R(t)\bm x_b$ that equals
$\bm v+\bm \Omega\times\bm x+\bm R\bm v_b(\bm R^t\bm x, t)$. This defines the no-slip boundary condition for the fluid flow $\bm u(\bm x-\bm x_0(t), t)$ that
we count from the center of the swimmer. Using the assumption that the Reynolds number is small, we find that
the flow $\bm u(\bm x, t)$ obeys unsteady Stokes equations,
\begin{eqnarray}&&\!\!\!\!\!\!\!
\partial_t\bm u
=-\nabla p+\nu \nabla^2 \bm u,\  \nabla\cdot\bm u=0,\ \ \bm u(\infty)=0, \label{crer}
\end{eqnarray}
where $p$ is pressure (minus the hydrostatic pressure) divided by the fluid density $\rho$, and $\nu$ is the fluid kinematic viscosity \cite{kimkar}. The boundary conditions are vanishing of the flow at infinity and,
\begin{eqnarray}&&\!\!\!\!\!\!\!
\bm u(|\bm x|\!=\!a)\!=\!\bm v\!+\!\bm \Omega\!\times\!\bm x\!+\!\bm u_b(\bm x, t),\ \ \bm u_b(\bm x, t)\!=\!\bm R\bm v_b(\bm R^t\bm x, t),
\nonumber
\end{eqnarray}
on the swimmer's surface. The surface flow $\bm u_b(\bm x, t)$ in the frame of the fluid is derived from $\bm v_b(\bm x, t)$ and is also tangential to the surface.

{\it Force and torque on the swimmer.} The flow determines the force $\bm F$ and the torque $\bm T$
applied on the swimmer by the fluid via surface integrals of the stress tensor $\bm \sigma$,
\begin{eqnarray}&&\!\!\!\!\!\!\!\!
\frac{\bm \sigma}{\rho}\!\!=\!\!-p\bm{I}\!+\!\nu(\nabla \bm u\!+\!
(\nabla \bm u)^t),\ \bm F\!=\!\!\!\int_S\!\! \bm \sigma \hat {\bm r}dS,\ \ 
\bm T\!=\!\!\!\int_S \!\!\bm r\!\times \!\bm \sigma \hat {\bm r}dS
,\nonumber
\end{eqnarray}
where $\bm{I}$ is identity matrix, the superscript $t$ stands for transpose, $dS$ is infinitesimal element of the swimmer's surface $S$  and $\hat {\bm r}$ is unit vector in radial direction.
We consider the swimmer as a ball having uniform mass density $\rho_s$ neglecting small density changes due to local deformations of the thin surface which represents short cilia or flagella.
Thus, the swimmer has mass $m_s=4\pi \rho_s a^3/3$ and constant moment of inertia $J=2 m_s a^2/5$. For simplicity, the effects of displacement of the center of mass from the sphere's center
(i. e. bottom-heaviness, see e. g. \cite{photo}), are not considered here. We have,
\begin{eqnarray}&&\!\!\!\!\!\!\!\!\!\!\!\!
m_s\frac{d\bm v}{dt}\!=\!\bm F+\!(m_s\!-\!m_F)\bm g,\ \ J\frac{d\bm \Omega}{dt}\!=\!\bm T, \label{fr}
\label{centm}
\end{eqnarray}
where $\bm g$ is the gravity acceleration and $m_F=4\pi a^3\rho/3$ is the fluid mass displaced by the sphere. We construct the flow which solves Eq.~(\ref{crer}) as superposition, $\bm u=\bm u^s+\bm u^r$, of the flow $\bm u^s$ for
inertialess swimmer in inertial fluid and the flow $\bm u^r$ for rigid sphere. The former obeys,
\begin{eqnarray}&&\!\!\!\!\!\!\!\!\!\!\!\!
\rho\partial_t\bm u^s=-\nabla p^s+\eta \nabla^2 \bm u^s,\  \nabla\cdot\bm u^s=0,\ \ \bm u^s(\infty)=0,\label{swimm}\\&&\!\!\!\!\!\!\!\!\!\!\!\!
\bm u^s(S)=\bm v^s(t)+\bm \Omega^s(t)\times\bm x+\bm u_b(\bm x, t),\ \ \bm F^s=\bm T^s=0.\nonumber
\end{eqnarray}
The flow $\bm u^r$ is the flow around a rigid sphere that moves with prescribed translational and angular velocities $\bm v-\bm v^s(t)$ and $\bm \Omega-\bm \Omega^s$, respectively. Thus, it obeys:
\begin{eqnarray}&&\!\!\!\!\!\!\!\!\!\!\!\!\!\!\!\!\!
\rho\partial_t\bm u^r=-\nabla p^r+\eta \nabla^2 \bm u^r,\  \nabla\cdot\bm u^r=0,\ \ \bm u^r(\infty)=0,\nonumber\\&&\!\!\!\!\!\!\!\!\!\!\!\!\!\!\!\!\!
\bm u^r(S)=\bm v(t)-\bm v^s(t)+\left[\bm \Omega(t)-\bm \Omega^s(t)\right]\times\bm x.
\end{eqnarray}
Since $\bm u^s$ imposes zero force and torque on the swimmer then $\bm F$ and $\bm T$ are determined by $\bm u^r$. The force $\bm F(t)$ is the force \cite{ll6,boussi} acting on a rigid sphere that moves with velocity $\bm v(t)-\bm v^s(t)$. This is given by Fourier representation,
\begin{eqnarray}&&\!\!\!\!\!\!\!\!\!\!\!
\bm F(t)\!=\!-6\pi\eta a\int_{-\infty}^{\infty}\!\! f(\delta\!-\!i\omega \tau_d) \left[\bm {\hat v}(\omega)\!-\!\bm {\hat v}^s(\omega)\right]e^{-i\omega t}d\omega,\nonumber\\&&\!\!\!\!\!\!\!\!\!\!\!
f(\lambda)=1+3\sqrt{\lambda}+\lambda,\ \ \tau_d=\frac{a^2}{9\nu},\label{f0}
\end{eqnarray}
where $\delta$ is infinitesimal so $\sqrt{\delta-i\omega}=\sqrt{|\omega|/2}(1-i \omega/|\omega|)$.
We designate Fourier transforms in time by hats. Similarly the torque $\bm T$ is given by \cite{ll6},
\begin{eqnarray}&&\!\!\!\!\!\!\!\!\!\!\!\!\!\!\!\!\!
\bm T(t)\!=\!-\int_{-\infty}^{\infty} T_0(\omega)\left[\bm {\hat  \Omega}(\omega)-\bm {\hat  \Omega}^s(\omega)\right]e^{-i\omega t}d\omega,
\nonumber\\&&\!\!\!\!\!\!\!\!\!\!\!\!\!\!\!\!\!
T_0(\omega)=\frac{5 J }{3\gamma \tau_d}\left(1 -\frac{3 i  \omega \tau_d}{1\!+\!q(\omega \tau_d)}\right)
,\ \  q(\omega)=3\sqrt{\delta-i\omega}, \label{t0}
\end{eqnarray}
where $\gamma=\rho_s/\rho$ is the specific gravity. Thus the force and the torque are determined by $\bm v^s$ and $\bm  \Omega^s$, respectively. This way of solution helps separating the effects brought by the inclusion of the inertia of the fluid and of the swimmer.

\section{Inertialess swimming in inertial fluid}

In this Section we derive translational and rotational velocities attained by an inertialess swimmer in inertial fluid. We circumvent finding $\bm u^s$ by using the reciprocal theorem \cite{maxeyriley,kimkar,stonesamuel,ardekani,gonzlauga},
\begin{eqnarray}&&\!\!\!\!\!\!\!\!\!\!\!\!\!\!\!\!\!
\bm {\hat v}^s\cdot\int_S\! \bm {\hat \sigma}^k\bm {\hat r}dS\!+\!a\epsilon_{irn} {\hat \Omega}^s_r\int_S \!{\hat \sigma}^k_{il}{\hat r}_l{\hat r}_ndS\!=\!-\!\!\int_S\! \bm {\hat u}_s\bm {\hat \sigma}^k \bm {\hat r}dS,\label{rcp}
\end{eqnarray}
where $\bm \sigma^k$ is the stress tensor of dual flow $\bm u^k$ also obeying unsteady Stokes equations. We use different dual flows for finding the translational and the rotational velocities.

\subsection{Translational velocity}

The dual flow, used for finding translational velocity of the swimmer, is the solution of unsteady Stokes equations whose Laplace transform obeys on the sphere ${\hat u}^k_i=\delta_{ik}$ and decays at infinity. In time domain the flow $\bm u^k$ describes motion of the sphere that starts from rest at $t=0$ and has impulsive velocity $u^k_i(|\bm x|=a, t)=\delta(t)\delta_{ik}$ (we do not keep dimensions here since the corresponding dimensional factors disappear from the final formulas). Thus at $t<0$ there is no flow, then the velocity jump at $t=0$ creates a flow. At $t>0$ the sphere is fixed at the origin and the flow caused by initial impulse decays. The Fourier transform is readily inferred from the Laplace transform provided in \cite{maxeyriley}, see also \cite{brg}. We have on the surface of the sphere,
\begin{eqnarray}&&\!\!\!\!\!\!\!\!\!\!\!\!\!\!\!\!\!
\frac{{\hat \sigma}^k_{il}{\hat r}_l}{3\eta}\!=\!\frac{ia\omega{\hat r}_i{\hat r}_k}{6\nu}\!-\!\frac{\delta_{ik}(1\!+\!q(\omega\tau_d) )}{2a},\label{surface}
\end{eqnarray}
where $q(\omega)$ is defined in Eq.~(\ref{t0}). We find using this flow in Eq.~(\ref{rcp}) that,
\begin{eqnarray}&&\!\!\!\!\!\!\!\!\!\!\!\!\!\!\!\!\!
\bm {\hat v}^s\!=\!\left(1\!+\!\frac{i\omega \tau_d }{f(\delta\!-\!i\omega \tau_d)}\right)\bm {\hat v}_0^s, \label{swtr0}
\end{eqnarray}
where $\bm v_0^s$ is the velocity of inertialess swimmer in inertialess fluid \cite{stonesamuel} which using definitions in Eq.~(\ref{crer}) reads,
\begin{eqnarray}&&\!\!\!\!\!\!\!\!\!\!\!\!\!\!\!\!\!
\bm v_0^s(t)=-\int_S \bm u_s(\bm x, t)\frac{dS}{4\pi a^2}=-\bm R
\int_S \bm v_b(\bm x, t)\frac{dS}{4\pi a^2}.
\label{swimming}
\end{eqnarray}
Inverse Fourier transform of Eq.~(\ref{swtr0}) gives,
\begin{eqnarray}&&\!\!\!\!\!\!\!\!\!\!\!\!\!\!\!\!\!
\bm v^s(t)=\bm v_0^s(t)-\frac{d}{dt}\int_{-\infty}^t K\left(\frac{t-t'}{\tau_d}\right)\bm v_0^s(t')dt', \label{infl}
\end{eqnarray}
where we introduced
\begin{eqnarray}&&\!\!\!\!\!\!\!\!\!\!\!\!\!\!\!\!\!
K(t)=\int \frac{\exp(-i\omega t)}{f(\delta\!-\!i\omega)}\frac{d\omega}{2\pi}.  \label{fou}
\end{eqnarray}
We used above that,
\begin{eqnarray}&&\!\!\!\!\!\!\!\!\!\!\!\!\!\!\!\!\!
\int \!\frac{i\omega \tau_d \bm {\hat v}_0^s \exp(-i\omega t)}{f(\delta\!-\!i\omega \tau_d)}\frac{d\omega}{2\pi}
\!=\!-\frac{d}{dt}\!\int\!\! K\left(\frac{t\!-\!t'}{\tau_d}\right)\bm v_0^s(t')dt'.
\end{eqnarray}
We demonstrate that the behavior of $K(t)$ is quite different from that of $t^{-1/2}$ memory kernel of the force on a rigid sphere \cite{Basset,boussi,maxeyriley,ll6}.

\subsection{Memory kernel}

We derive the memory kernel $K(t)$ in Eq.~(\ref{infl}). We observe that $f^{-1}(z)$ is analytic in the complex plane with branch cut at negative real semi-axis. Consequently $f^{-1}(\delta-i\omega)$ considered as function of complex variable $\omega$ is analytic outside the branch cut at $(-i\infty, -i\delta)$. We find $K(t)=0$ for $t<0$ because the integration contour can be closed in the upper half plane producing zero. This is necessary for causality: otherwise instantaneous velocity in Eq.~(\ref{infl}) would be determined by future movements of the swimmer. When $t>0$ we find passing in Eq.~(\ref{fou}) to the integration variable $\lambda=\delta-i\omega$ that,
\begin{eqnarray}&&\!\!\!\!\!\!\!\!\!\!\!\!\!\!\!\!\!
K(t)=\int \frac{\exp(-i\omega t)}{f(\delta\!-\!i\omega)}\frac{d\omega}{2\pi}=\int_{\delta-i\infty}^{\delta+i\infty} \frac{\exp(\lambda t)}{f(\lambda)}\frac{d\lambda}{2\pi i},
\end{eqnarray}
that is, $K(t)$ is the inverse Laplace transform of $1/f(\lambda)$. This conclusion could be reached directly using the Laplace transform instead of the Fourier transform in the derivations. However this change would have other disadvantages. For use in the next Section we consider the slightly more general inverse Laplace transform
\begin{eqnarray}&&\!\!\!\!\!\!\!\!\!\!\!\!\!\!
{\tilde K}(t)\!\!=\!\!\!\int_{\delta-i\infty}^{\delta+i\infty}\!\! \frac{\exp(\lambda t)}{{\tilde f}(\lambda)}\frac{d\lambda}{2\pi i},\ \  {\tilde f}(\lambda)=1+3\sqrt{\lambda}+\kappa \lambda. \label{tld11}
\end{eqnarray}
The kernel $K(t)$ is obtained from ${\tilde K}(t)$ by setting $\kappa=1$. We use that,
\begin{eqnarray}&&
\frac{1}{1+3\sqrt{\lambda}+\kappa \lambda}= \frac{1}{\sqrt{9-4\kappa}}\left[\frac{x_2}{\lambda+x_2\sqrt{\lambda}}-\frac{x_1}{\lambda+x_1\sqrt{\lambda}}\right],\nonumber
\end{eqnarray}
where $-x_i$ are roots of quadratic polynomial $1+3x+\kappa x^2$,
\begin{eqnarray}&&
x_1=\frac{3-\sqrt{9-4\kappa}}{2\kappa},\ \
x_2=\frac{3+\sqrt{9-4\kappa}}{2\kappa}.
\end{eqnarray}
We use known integral (cf. similar calculation in \cite{ishimoto}),
\begin{eqnarray}&&\!\!\!\!\!\!\!\!\!\!\!\!\!\!\!\!\!
\int_0^{\infty}\!\!\!\exp\left(-\lambda t+x_i^2t\right) {\rm Erfc}(x_i\sqrt{t})dt=\frac{1}{\lambda+x_i\sqrt{\lambda}},\nonumber\end{eqnarray}
where analytic continuation is used for defining the error function ${\rm Erfc}(x)$ for complex $x$. We conclude that,
\begin{eqnarray}&&
{\tilde K}(t)=\frac{1}{\sqrt{9-4\kappa}}\left[x_2\exp\left(x_2^2t\right){\rm Erfc}(x_2\sqrt{t})\right.\nonumber\\&&\left.
-x_1\exp\left(x_1^2t\right){\rm Erfc}(x_1\sqrt{t})\right].
\end{eqnarray}
This can be represented as series,
\begin{eqnarray}&&\!\!\!\!\!\!\!\!\!
{\tilde K}(t)=\frac{1}{\sqrt{9-4\kappa}}\left[x_2\exp\left(x_2^2t\right)-x_1\exp\left(x_1^2t\right)\right.\nonumber\\&&\!\!\!\!\!\!\!\!\!\left.
+\frac{2x_1^2\sqrt{t}}{\sqrt{\pi}}\sum_{k=0}^{\infty}\frac{(2t)^k x_1^{2k}}{(2k+1)!!}-\frac{2x_2^2\sqrt{t}}{\sqrt{\pi}}\sum_{k=0}^{\infty}\frac{(2t)^k x_2^{2k}}{(2k+1)!!}\right],
\end{eqnarray}
which is useful at small $t$. In contrast to the memory kernel for the force on the rigid sphere, that has square root divergence at zero, ${\tilde K}(t)$ has a finite value of $1/\kappa$ at $t=0$. When $t$ is large we can use,
\begin{eqnarray}&&
{\tilde K}(t)=\frac{1}{\pi\sqrt{9-4\kappa}}\left[\sum_{k=0}^{\infty}\frac{(-1)^k\Gamma(k+3/2)}{t^{k+3/2}x_1^{2(k+1)}}\right.\nonumber\\&&\left.
-\sum_{k=0}^{\infty}\frac{(-1)^k\Gamma(k+3/2)}{t^{k+3/2}x_2^{2(k+1)}}\right].
\end{eqnarray}
The leading order behavior at large times is given by $k=0$ term. We find using $x_1^{-2}-x_2^{-2}=3\sqrt{9-4\kappa}$ that ${\tilde K}(t)\sim 3t^{-3/2}/\sqrt{4\pi}$. This behavior is independent of $\kappa$ because the leading order behavior of ${\tilde f}(\lambda)$ at small $\lambda$, given by $[1+3\sqrt{\lambda}]^{-1}$, is independent of $\kappa$. Thus it holds also that $K(t)\sim 3t^{-3/2}/\sqrt{4\pi}$.

In the case of Volvox the density of the swimmer is approximately the density of the fluid, $\kappa\approx 3$, so that $\sqrt{9-4\kappa}\approx i\sqrt{3}$ and $x_i$ are complex conjugates of each other. We find,
\begin{eqnarray}&&\!\!\!\!\!\!\!\!\!\!\!\!\!\!\!\!
{\tilde K}(t)\!=\!\frac{2{\cal I}\left[ x_2\exp\left(x_2^2t\right){\rm Erfc}(x_2\sqrt{t})\right]}{\sqrt{3}} ,\ \  x_2\!=\!\frac{3\!+\!i\sqrt{3}}{6}.
\end{eqnarray}
Finally, we find $K(t)$ setting $\kappa=1$ above which gives,
\begin{eqnarray}&&
x_2=\frac{3+\sqrt{5}}{2},\ \ x_1=\frac{3-\sqrt{5}}{2},\nonumber\\&&
K(t)=\frac{1}{\sqrt{5}}\left[x_2\exp\left(x_2^2t\right){\rm Erfc}(x_2\sqrt{t})\right.\nonumber\\&&\left.
-x_1\exp\left(x_1^2t\right){\rm Erfc}(x_1\sqrt{t})\right],
\end{eqnarray}
with the corresponding asymptotic forms and series. We see that, in contrast with the $t^{-1/2}$ memory kernel for the force that diverges at zero, $K(0)$ is finite and equals one. Moreover, $K(t)$ is integrable
due to $K(t)\sim 3t^{-3/2}/\sqrt{4\pi}$ behavior at large times. We have $\int_0^{\infty} K(t)dt=1/f(\lambda=0)=1$. Thus $K(t)$ is similar to a delta-function smeared over a scale of order one. The integral in Eq.~(\ref{infl}) is determined by $|t-t'|\lesssim \tau_d$ where $\tau_d$ is defined in Eq.~(\ref{f0}). If $Ro\equiv \sigma\tau_d\ll 1$ then $\tau_d$ is much smaller than the swimming period $2\pi/\sigma$ and we can set $\bm v_0^s(t')\approx \bm v_0^s(t)$ in the integrand. We find the leading order correction in the fluid inertia $\bm v^s=\bm v_0^s-\tau_d d\bm v_0^s/dt$. The correction is small which is consistent with the negligibility of time derivative term in the NSE at $Ro\ll 1$.

\subsection{Rotational velocity}

The dual flow which we use in Eq.~(\ref{rcp}) for finding rotational velocity of the swimmer is the solution of unsteady Stokes equations whose Laplace transform obeys on the sphere ${\hat u}^k_i=\epsilon_{ikn} x_n$ and decays at infinity. This is flow which is created by instantaneous rotation of the sphere at $t=0$ with angular velocity given by unit vector in $k-$th direction: $u^k_i(x=a, t)=\epsilon_{ikn} x_n\delta(t)$. This flow can be obtained as superposition of flows caused by rotation of the sphere at angular velocity that depends on time as $\exp[-i\omega t]$ found in \cite{ll6}. We find ($y=|\bm x|/a$ and $q=q(\omega\tau_d)$),
\begin{eqnarray}&&\!\!\!\!\!\!\!\!\!\!\!\!\!\!\!\!
u^k_i(\bm x, t)\!=\!\frac{\epsilon_{ikn}x_n}{y^3}\int\!\! \frac{d\omega}{2\pi}\exp\left[-i\omega t\!-\!q(y\!-\!1)\right]\frac{1\!+\!qy}{1\!+\!q}.
\end{eqnarray}
If we use spherical coordinates with polar axis in $k-$th direction then velocity has the only non-vanishing component $u_{\phi}=v$ where \cite{ll6},
\begin{eqnarray}&&\!\!\!\!\!\!\!\!\!\!\!\!\!\!\!\!\!
v=\frac{x\sin\theta}{y^3} \int \frac{d\omega}{2\pi}\exp\left[-i\omega t-q(y-1)\right]\frac{1+qy}{1+q}.
\end{eqnarray}
The only non-zero component of $\sigma^k_{il}{\hat x}_l$ on the surface $S$ is \cite{ll6},
\begin{eqnarray}&&\!\!\!\!\!\!\!\!
\frac{\sigma^k_{\phi r}}{\eta}\!=\!\left(\frac{\partial v}{\partial x}\!-\!\frac{v}{x}\right)|_{S}\!=\!-\sin\theta\left(3\delta(t)\!+\!\int \frac{q^2d\omega}{2\pi(1+q)}e^{-i\omega t}\right),\nonumber\\&&\!\!\!\!\!\!\!\!
\frac{{\hat \sigma}^k_{il}{\hat x}_l}{\eta}=-\epsilon_{ikl}{\hat x}_l\left[3+\frac{q^2}{1+q}\right].
\end{eqnarray}
where the last line is written in the form independent of the reference frame. We find using this in Eq.~(\ref{rcp}) that,
\begin{eqnarray}&&\!\!\!\!\!\!\!\!\!\!\!\!\!\!\!\!\!
\bm  \Omega^s(t)\!=\!-\!\!\int_S\!\!  \frac{3\bm x\! \times\! \bm u_s(\bm x, t)dS}{8\pi a^4}\! =\!-\bm R\int_S\!\!\frac{3\bm x\!\times\! \bm v_b(\bm x, t) dS}{8\pi a^4}. \label{rt}
\end{eqnarray}
This formula is identical to the formula for rotational swimming velocity at zero inertia \cite{stonesamuel} which has a remarkable consequence, as explained below.

\subsection{Irrelevance of fluid inertia for mean velocity and scallop theorem}

The above formulas imply that the mean swimming velocity of inertialess swimmer in inertial fluid does not differ at all from the
mean velocity of inertialess swimmer in inertialess fluid. Indeed, we observe that the last term in Eq.~(\ref{infl}) is time derivative of a bounded function. Thus it gives no contribution in time averaged velocity. In other words, time-averaged velocity obeys,
\begin{eqnarray}&&\!\!\!\!\!\!\!\!\!\!\!\!\!\!\!\!\!
\langle \bm v^s\rangle\!=\!-\frac{1}{4\pi a^2}\left\langle \bm R
\int_S \bm v_b(\bm x, t) dS\right\rangle, \label{dcgsdu}
\end{eqnarray}
where the angular brackets stand for time average, cf. with $\bm {\hat v}^s(\omega=0)=\bm {\hat v}_0^s(\omega=0)$ in Eq.~(\ref{swtr0}). However, the angular velocity is the same as at zero fluid inertia, see Eq.~(\ref{rt}). Thus $\bm R$ and surface-averaged $-\bm R \bm v_b$ for inertial fluid do not differ from those for inertialess fluid. The mean swimming velocity is independent of the fluid inertia as long as $\bm v_b$ can be considered as given.

We conclude that the fluid inertia adds only oscillatory motions that bring no net propulsion but can be relevant for nutrient uptake \cite{ardekani}. Correspondingly the
scallop theorem stating that net propulsion is impossible for time-reversible strokes when neglecting both inertia of the fluid and of the swimmer \cite{purcell} can be extended to include the fluid inertia. This is in agreement with
concrete calculations of \cite{ardekani,gonzlauga,ishimoto}.

\section{Including inertia of the swimmer}

We now also incorporate the inertia of the swimmer. We found $\bm v^s$ and thus we can write the force in Eq.~(\ref{centm}) which would reproduce the equation of motion derived in \cite{ardekani}.
The derivation here provides a different insight in the structure of the flow around the swimmer. Our use of this equation is completely different from the use made in \cite{ardekani} described previously.

\subsection{Translational velocity with inertia}

It is simpler to find the solution directly in Fourier space.
Performing Fourier transform of Eq.~(\ref{fr}) and using Eq.~(\ref{f0}) we obtain
\begin{eqnarray}&&\!\!\!\!\!\!\!\!
-i\omega \bm {\hat v}\!=\!-\frac{f(\delta\!-\!i\omega \tau_d)\left[\bm {\hat v}\!-\!\bm {\hat v}^s\right]}{2\gamma\tau_d}\!+\!\frac{2\pi\delta(\omega)(\gamma-1)\bm g}{\gamma} 
,\end{eqnarray}
where 
we used that initial conditions on velocity in the remote past are forgotten. We obtain the swimmer's velocity by solving for $\bm {\hat v}$ and
introducing $\kappa=1+2\gamma$,
\begin{eqnarray}&&\!\!\!\!\!\!\!\!
\bm {\hat v}(\omega)\!=\!\left(1+\frac{i\kappa \omega \tau_d}{{\tilde f}(\delta\!-\!i\omega \tau_d)}\right)\bm {\hat v}_0^s\!+\!4\pi\delta(\omega)(\gamma-1)\bm g\tau_d  \label{for}
,\end{eqnarray}
where ${\tilde f}$ is defined in Eq.~(\ref{tld11}) and the last term is sedimentation velocity. We find using similarity with Eqs.~(\ref{swtr0})-(\ref{infl}),
\begin{eqnarray}&&\!\!\!\!\!\!\!\!\!\!\!\!\!\!\!\!\!
\bm v(t)=\bm v_0^s(t)-\kappa\frac{d\bm x_{i}(t)}{dt}+\!2(\gamma-1)\bm g\tau_d,\label{solutioni}
\end{eqnarray}
where we introduced the inertial displacement $\bm x_{i}(t)$,
\begin{eqnarray}&&\!\!\!\!\!\!\!\!\!\!\!\!\!\!
\bm x_i=\int_{-\infty}^t{\tilde K}\left(\frac{t-t'}{\tau_d}\right)\bm v_0^s(t')dt', \label{tld}
\end{eqnarray}
with ${\tilde K}$ from Eq.~(\ref{tld11}). The propulsion velocity given by Eqs.~(\ref{solutioni})-(\ref{tld}) reduces to that for inertialess swimmer in Eqs.~(\ref{swtr0})-(\ref{infl}) by taking $\gamma\to 0$ and $\kappa\to 1$. This is because $K(t)$ is ${\tilde K}(t)$ at $\kappa=1$. For Volvox $\gamma\approx 1$, see \cite{Goldstein}, and $\kappa\approx 3$ so inertia of the swimmer brings a finite change of the swimmer's velocity.

We demonstrated in the previous Section that ${\tilde K}(t)$ is roughly a delta-function smeared over $t\sim 1$. Thus we find by performing consideration similar to that after Eq.~(\ref{infl}),
\begin{eqnarray}&&\!\!\!\!\!\!\!\!\!\!
\bm v\approx \bm v_0^s-\kappa\tau_d \frac{d\bm v_0^s}{dt}, \ \ \ \ Ro\ll 1.  \label{trs}
\end{eqnarray}
The inertial displacement $\bm x_{in}(t)$ is bounded and does not contribute the time averaged velocity,
\begin{eqnarray}&&\!\!\!\!\!\!\!\!\!\!\!\!\!\!\!\!\!
\langle \bm v\rangle\!=\!-\frac{1}{4\pi a^2}\left\langle \bm R
\int_S \bm v_b(\bm x, t) dS\right\rangle\!+2(\gamma-1)\bm g\tau_d.\label{d}
\end{eqnarray}
where we used Eqs.~(\ref{swimming}) and (\ref{solutioni}). This has the same form as the average velocity of inertialess swimmer in inertialess fluid \cite{stonesamuel}. This also agrees with the average velocity of inertialess swimmer in inertialess fluid given by Eq.~(\ref{dcgsdu}) with sedimentation velocity included. Thus inertia can bring a difference only by changing the rotation matrix $\bm R$. This has the useful consequence, as explained below.

\subsection{Irrelevance of inertia for mean speed at axial symmetry}

The case of axially symmetric squirmers is much studied and presents separate interest. In this case the vectors of translational and angular velocities are parallel
so the swimmer propagates as a screw, see \cite{swirl} for an example. Then $\bm R$ has no effect in Eq.~(\ref{d}) and can be dropped. The classical swimming theory neglecting inertia applies to the average velocity. We conclude that inertial axially symmetric tangential squirmers obey the scallop theorem as was observed in \cite{gonzlauga,ardekani,ishimoto}.

\subsection{Metachronal wave}

We consider the metachronal wave as an example of axially symmetric swimming. This stroke is believed to describe Volvox with good approximation \cite{swirl}. Similarly to \cite{stonesamuel}, we consider only tangential part described by time $t$ position of polar angle $\theta$ as a function of the position $\theta_0$ near which the oscillation occurs as $\theta(t, \theta_0)=\!\theta_0\!+\!\epsilon \cos(k\theta_0\!-\!\sigma t)$. Our consideration disregards radial and azimuthal displacements. The parameters that fit experiment are $k=4.7$, $\sigma=203$ radian per second and $\epsilon\approx 0.06$, see details in \cite{swirl}. For this stroke there is no rotation and the stroke has only $\theta-$component $v_{b\theta}(t, \theta=\theta(t, \theta_0))=a\partial_t\theta(t, \theta_0)$, cf. Eq.~(\ref{impl}). We have to order $\epsilon$ that $\theta_0=\theta(t, \theta_0)-\epsilon\cos(k\theta(t, \theta_0)\!-\!\sigma t)$ which on taking the derivative gives,
\begin{eqnarray}&&\!\!\!\!\!\!\!\!\!\!
v_{b\theta}(t, \theta)
\approx a\sigma \epsilon \sin\left(k\theta\!-\!\sigma t\!-\!k\epsilon \cos(k\theta\!-\!\sigma t)\right),
\end{eqnarray}
which neglects terms of order $\epsilon^3$ and higher. Expanding the last line in the same order,
\begin{eqnarray}&&\!\!\!\!\!\!\!\!\!\!\!\!\!\!\!
v_{b\theta}(t, \theta, \phi)\approx a\sigma\epsilon \sin\left(k\theta\!-\!\sigma t\right)-
ak\sigma\epsilon^2 \cos^2\left(k\theta\!-\!\sigma t\right).\label{tang}
\end{eqnarray}
The axial symmetry dictates that $\bm v_0^s(t)$ points in vertical direction. We obtain designating this component by $v_0^s(t)$ and using $v_{bz}=-v_{b\theta}\sin\theta$ that,
\begin{eqnarray}&&\!\!\!\!\!\!\!\!\!\!\!
v_0^s(t)=\frac{1}{2}\int_0^{\pi} v_{b\theta}\sin^2\theta d\theta=\frac{1}{4}\int_0^{\pi} v_{b\theta}(1-\cos(2\theta)) d\theta.\nonumber
\end{eqnarray}
We find using Eq.~(\ref{tang}) and the integrals
\begin{eqnarray}&&\!\!\!\!\!\!\!\!\!\!\!\!\!\!\!\!\!
\int_0^{\pi} \sin(k\theta\!-\!\sigma t)\sin^2\theta d\theta=\frac{4\sin(\pi k/2)\sin(\pi k/2-\sigma t)}{k(4-k^2)}.
\nonumber\\&&\!\!\!\!\!\!\!\!\!\!\!\!\!\!\!\!\!
\int_0^{\pi}\!\! \cos^2(k\theta\!-\!\sigma t)\sin^2\theta d\theta\!=\!\frac{\pi}{4}\!+\!\frac{\sin(\pi k)\cos(\pi k\!-\!2\sigma t)}{4k(1\!-\!k^2)}.\nonumber
\end{eqnarray}
that the swimming velocity neglecting inertia obeys,
\begin{eqnarray}&&\!\!\!\!\!\!\!\!
v_0^s(t)=a\sigma\epsilon \left(
\frac{2\sin(\pi k/2)\sin(\pi k/2-\sigma t)}{k(4-k^2)}-\frac{\pi k\epsilon }{8}
\right.\nonumber\\&&\!\!\!\!\!\!\!\!\left.
-\frac{\epsilon \sin(\pi k)\cos(\pi k-2\sigma t)}{8(1-k^2)}\right), \label{mt}
\end{eqnarray}
up to order $\epsilon^2$. The time-average is determined by the only term which does not oscillate $\langle v_0^s(t)\rangle=-\pi a\sigma k\epsilon^2/8$ reproducing result of \cite{stonesamuel}.

We now consider the inertial corrections. We have by Fourier transform of Eq.~(\ref{mt}),
\begin{eqnarray}&&\!\!\!\!\!\!\!\!
\frac{{\hat v}_0^s}{a\sigma\epsilon{\tilde f}(\delta\!-\!i\omega \tau_d)}\!=\!
\frac{2i \pi  \sin(\pi k/2)}{k(4\!-\!k^2)}\left(\frac{ \exp(\!-i \pi k/2)\delta(\omega\!+\!\sigma)}{{\tilde f}(\delta\!+\!i\sigma \tau_d)}\right.\nonumber\\&&\!\!\!\!\!\!\!\!\left.
-\frac{\exp(i \pi k/2)\delta(\omega-\sigma)}{{\tilde f}(\delta-i\sigma \tau_d)}\right)-\frac{\pi \epsilon\sin(\pi k)}{8(1-k^2)}\nonumber\\&&\! \!\!\!\!\!\!\! \cdot \left(
\frac{\exp(i \pi k/2)\delta(\omega-\sigma)}{{\tilde f}(\delta-i\sigma \tau_d)}
+\frac{ \exp(-i \pi k/2)\delta(\omega+\sigma)}{{\tilde f}(\delta+i\sigma \tau_d)}\right),\label{velp}
\end{eqnarray}
where we did not write the $\delta(\omega)$ term which does not contribute to the correction. We observe that,
\begin{eqnarray}&&\!\!\!\!\!\!\!\!
{\tilde f}(\delta\pm i\sigma \tau_d)=1\pm \frac{i\kappa Ro}{9} +\sqrt{\frac{Ro}{2}}(1\pm i),
\end{eqnarray}
so that ${\tilde f}(\delta+ i\sigma \tau_d)$ and ${\tilde f}(\delta- i\sigma \tau_d)$ are complex conjugates. Using Eq.~(\ref{for}) and performing inverse Fourier transform of Eq.~(\ref{velp}), we find
the formula for velocity including inertia of both the fluid and the swimmer,
\begin{eqnarray}&&\!\!\!\!\!\!\!\!
\bm v(t)=\bm v_0^s(t)+\!2(\gamma-1)\bm g\tau_d-a\epsilon \kappa Ro\frac{d}{dt}\left(
\frac{2  \sin(\pi k/2)}{k(4-k^2)}
\right.\label{wave}
\\&&\!\!\!\!\!\!\!\!\left.\cdot
{\cal I}\left[\frac{\exp(i \pi k/2\!-\!i\sigma t)}{{\tilde f}(\delta\!-\!i Ro)}\right]
\!-\!{\cal R}\left[\frac{\epsilon\sin(\pi k)\exp(i \pi k/2\!-\!i\sigma t)}{8(1-k^2){\tilde f}(\delta-i Ro)}\right]\right),
\nonumber\end{eqnarray}
where ${\cal I}$ and ${\cal R}$ stand for imaginary and real part, respectively. It can be concluded that inertia does not change the time-average value of $\bm v(t)$ for this stroke, whereas oscillations of $\bm v(t)$ contain an inertial correction, cf. \cite{ardekani}.

\subsection{Rotational velocity with inertia}

Rotational velocity of inertial swimmer obeys $J\dot{\bm \Omega}=\bm T$. The Fourier transform gives,
\begin{eqnarray}&&\!\!\!\!\!\!\!\!\!\!\!
\bm {\hat \Omega}(\omega)=\bm {\hat \Omega}^s(\omega)+\frac{iJ\omega \bm {\hat \Omega}^s(\omega)}{T_0(\omega)-iJ\omega}=\frac{\bm {\hat \Omega}^s(\omega)}{1-iJ\omega  T_0^{-1}(\omega)},
\label{rp} 
\end{eqnarray}
where we used Eq.~(\ref{t0}). This solves for the swimmer's rotation implicitly with $\bm  \Omega^s$ from Eq.~(\ref{rt}).
In time domain,
\begin{eqnarray}&&\!\!\!\!\!\!\!\!\!\!\!
\bm \Omega(t)=\bm \Omega^s(t)-\frac{d}{dt}\int_{-\infty}^t K_r\left(\frac{t-t'}{\tau_d}\right)\bm \Omega^s(t')dt',\label{angvl}
\end{eqnarray}
where using Eq.~(\ref{t0}) we introduced the kernel,
\begin{eqnarray}&&\!\!\!\!\!\!\!\!\!\!\!\!\!\!
K_r(t)=\int_{-\infty}^{\infty}\frac{d\omega}{2\pi} \frac{\exp(-i\omega  t)}{5\left(1 -3 i  \omega/[1\!+\!q(\omega)]\right)/\left(3\gamma\right)-i\omega}.
\end{eqnarray}
We assume that $5/(3\gamma)\sim 1$ so that this equation gives that the characteristic time of variations of $K_r(t)$ is $O(1)$.
We observe that $\int_0^{\infty} K_r(t)dt$ is finite and given by the Fourier transform at zero frequency which is $3\gamma/5$. We conclude that $K_r(t)$ decays over time of order one. In the limit of small Roshko number in the leading order we can set $\bm \Omega^s(t')\approx \bm \Omega^s(t)$ in the integrand finding,
\begin{eqnarray}&&\!\!\!\!\!\!\!\!\!\!\!
\bm \Omega(t)\approx \bm \Omega^s(t)-\frac{3\tau_s}{10}\frac{d\bm \Omega^s(t)}{dt},\ \ Ro\ll 1. \label{tos}
\end{eqnarray}
where $\tau_s=2\gamma\tau_d$ is the Stokes time. This formula describes smaller organisms.

\subsection{Inertia changes mean swimming speed via translation-rotation coupling}

Inertia of the swimmer, in contrast with the fluid inertia, changes the rotation of the swimmer. Thus, generally, inertia changes $\bm R$ and the net propulsion velocity. This translational-rotational coupling can bring qualitative and quantitative changes in the swimming.

The changes can be considered using Eqs.~(\ref{solutioni}) and (\ref{angvl}) describing the propagation of inertial swimmer in the inertial fluid. Rotation decouples from translation and can be considered separately. In contrast, translation depends on rotation via the rotation matrix $\bm R$ obeying $\dot {R}_{ij}=\epsilon_{ikl}\Omega_kR_{lj}$, see e. g. Eq.~(\ref{d}). The system of Eqs.~(\ref{solutioni}) and (\ref{angvl}) describing the translation-rotation coupling is non-local in time having memory described by the kernels ${\tilde K}(t)$ and $K_r(t)$. It becomes local in the limit of small inertia, $Ro\ll 1$, see Eqs.~(\ref{trs}) and (\ref{tos}).
Other limit where the equations becomes local is that of heavy swimmers whose density is much larger than that of the fluid, $\gamma\gg 1$, as explained below.

\subsection{Heavy swimmers} \label{sw}

We consider the limit of heavy (called dense in \cite{gonzlauga}) swimmers whose density is much larger than the density of the fluid, $\gamma\gg 1$. In this limit $\kappa\approx 2\gamma \to\infty$ and we can neglect $3\sqrt{\lambda}$ in ${\tilde f}(\lambda)=1+3\sqrt{\lambda}+\kappa \lambda$ (we observe that we cannot neglect the first term in ${\tilde f}(\lambda)$ that is relevant at $\lambda\lesssim 1/\kappa$). Thus Eq.~(\ref{for}) becomes
\begin{eqnarray}&&\!\!\!\!\!\!\!\!
\bm {\hat v}(\omega)\!=\!\frac{\bm {\hat v}_0^s}{1-i\omega \tau_s}\!+\!2\pi\delta(\omega)\bm g\tau_s,
\end{eqnarray}
where we observed that in this limit $\kappa\tau_d$ is the Stokes time $\tau_s=2\gamma\tau_d$. Inverse Fourier transform gives,
\begin{eqnarray}&&\!\!\!\!\!\!\!\!\!\!\!\!\!\!\!\!\!
\frac{d\bm v}{dt}=-\frac{\bm v-\bm  v_0^s}{\tau_s}+\bm g. \label{dense}
\end{eqnarray}
This reproduces the equation for heavy spherical swimmers derived in \cite{gonzlauga}. The limit of small inertia, $\sigma \tau_s\ll 1$, is found by writing $\bm v=\bm  v_0^s-\tau_s\dot{\bm v}+\bm g\tau_s$ and making first iteration of the RHS,
\begin{eqnarray}&&\!\!\!\!\!\!\!\!\!\!\!\!\!\!\!\!\!
\bm v \approx \bm  v_0^s-\tau_s\frac{d\bm v_0^s}{dt}+\bm g\tau_s.
\end{eqnarray}
This agrees with Eq.~(\ref{trs}). Similar consideration holds for the angular velocity. We have from Eq.~(\ref{rp})
\begin{eqnarray}&&\!\!\!\!\!\!\!\!\!\!\!
\bm {\hat \Omega}(\omega)=\frac{\bm {\hat \Omega}^s(\omega)}{1-(3i\omega\gamma \tau_d /5)  \left(1 -3 i  \omega \tau_d/[1\!+\!q(\omega \tau_d)]\right)^{-1}},
\end{eqnarray}
where we used the definition of $T_0(\omega)$ in Eq.~(\ref{t0}). We observe that in the limit of large $\gamma$ the characteristic frequency $\omega_c$ that defines the inverse Fourier transform of $\bm {\hat \Omega}(\omega)$ obeys $\omega_c\tau_d \sim \gamma^{-1}$. This is obtained by the demand that the prefactor in the denominator is of order one. We find observing that at these frequencies $\left(1 -3 i  \omega \tau_d/[1\!+\!q(\omega \tau_d)]\right)^{-1}\approx 1$ that,
\begin{eqnarray}&&\!\!\!\!\!\!\!\!\!\!\!
\bm {\hat \Omega}(\omega)\approx \frac{\bm {\hat \Omega}^s(\omega)}{1-3i\omega\tau_s /10},
\end{eqnarray}
where we used $2\gamma\tau_d=\tau_s$. Performing inverse Fourier transform,
\begin{eqnarray}&&\!\!\!\!\!\!\!\!\!\!\!\!\!\!\!\!\!
\frac{d\bm \Omega}{dt}=-\frac{10\left(\bm \Omega-\bm  \Omega^s\right)}{3\tau_s}.
\end{eqnarray}
We find combining the above the system of equations on coupled translational and rotational degrees of freedom,
\begin{eqnarray}&&\!\!\!\!\!\!\!\!\!\!\!
\tau_s\frac{d\bm v}{dt}\!+\!\bm v\!=\!-\frac{1}{4\pi a^2}\bm R \int\!\! \bm v_b(\bm x, t) dS, \ \ \frac{dR_{il}}{dt}\!=\!\epsilon_{ins}\Omega_nR_{sl},\nonumber \\&&\!\!\!\!\!\!\!\!\!\!\!
\frac{3\tau_s}{10}\frac{d\bm \Omega}{dt}+\bm \Omega=-\frac{3}{8\pi a^4}\bm R \int \bm x\times \bm v_b(\bm x, t) dS.\label{heavy}
\end{eqnarray}
This system was obtained in \cite{gonzlauga} from steady Stokes equations. Setting $\tau_s=0$ recovers the often-used formulation of Stone and Samuel \cite{stonesamuel} where inertia of the fluid and the swimmer is neglected \cite{stonesamuel}. In this case reciprocal theorem holds telling that if the swimming stroke is time-reversible then no net self-propulsion occurs over the stroke's period. The limit of small inertia, $\sigma \tau_s\ll 1$, for rotational velocity is found by writing $\bm \Omega=\bm \Omega^s-3\tau_s\dot{\bm \Omega}/10$ and making first iteration of the RHS. This reproduces Eq.~(\ref{tos}).

\section{Translational-rotational coupling at small inertia} \label{ro}

In this Section, we analyze the previously derived equations of motion at small Roshko number. We demonstrate that the leading order correction in $Ro$ to the mean swimming velocity is proportional to the cross product of translational and rotational swimming velocities at $Ro=0$. Thus the swimming velocity is given via the swimming stroke $\bm v_b(\bm x, t)$. Our starting point are Eqs.~(\ref{trs}) and (\ref{tos}),
\begin{eqnarray}&&\!\!\!\!\!\!\!\!
\bm v=\bm v_0^s-\kappa\tau_d \frac{d\bm v_0^s}{dt},\ \ \bm \Omega(t)=\bm \Omega^s(t)-\frac{3\tau_s}{10}\frac{d\bm \Omega^s(t)}{dt}. \label{smro}
\end{eqnarray}
These velocities depend on inertia directly, through the corrections linear in $Ro\propto \tau_d$, and indirectly through the rotation matrix, see Eq.~(\ref{swimming}). We use for studying the rotation matrix that,
\begin{eqnarray}&&\!\!\!\!\!\!\!\!\!\!\!\!
\frac{dR_{ij}}{dt}=R_{il}\epsilon_{lpj}\Omega^b_p,\ \ \bm \Omega^b=\bm R^t \bm \Omega(t),\label{rotationale}
\end{eqnarray}
where we introduced angular velocity in the body-fixed frame $\bm \Omega^b$ with $\bm R^t$ the matrix transpose of $\bm R$. This equation can be obtained from $\dot {R}_{ij}=\epsilon_{ikl}\Omega_kR_{lj}$
using rotational invariance of the Levi-Civita tensor that implies that $\epsilon_{tpj}=\epsilon_{ikl}R_{it}R_{kp}R_{lj}$ for any orthogonal $\bm R$. Indeed, this identity gives $R_{qt}\epsilon_{tpj}=\epsilon_{qkl}R_{kp}R_{lj}$, and using this in $\dot {R}_{ij}=\epsilon_{ikl}R_{kp}R_{lj}\Omega^b_p$ brings Eq.~(\ref{rotationale}). We find using Eqs.~(\ref{rt}) and (\ref{smro}) that in linear order in $Ro$,
\begin{eqnarray}&&\!\!\!\!\!\!\!\!
\bm \Omega^b=-\int_S\!\!\frac{3\bm r\!\times\! \bm v_b(\bm r, t) dS}{8\pi a^4}+\frac{3\gamma \tau_d}{5}\frac{d}{dt}\int_S\!\!\frac{3\bm r\!\times\! \bm v_b(\bm r, t) dS}{8\pi a^4}
\nonumber\\&&\!\!\!\!\!\!\!\!
+\frac{3\gamma \tau_d}{5}\bm \Omega^b\times \left( \int_S\!\!\frac{3\bm r\!\times\! \bm v_b(\bm r, t) dS}{8\pi a^4}\right),
\end{eqnarray}
where we used $\bm R^t \left(\bm \Omega\times \bm \Omega^s\right)=\bm \Omega^b\times \left(\bm R^t \bm \Omega^s\right)$. Iterating this equation (that contains $\bm \Omega^b$ on both sides) we find that in linear order in $Ro$ the last term can be dropped,
\begin{eqnarray}&&\!\!\!\!\!\!\!\!\!\!\!\!\!\!
\bm \Omega^b\!=\!-\!\int_S\!\!\frac{3\bm r\!\times\! \bm v_b(\bm r, t) dS}{8\pi a^4}\!+\!\frac{3\tau_s}{10}\frac{d}{dt}\!\int_S\!\!\frac{3\bm r\!\times\! \bm v_b(\bm r, t) dS}{8\pi a^4}.
\end{eqnarray}
Thus, $\bm \Omega^b$, in contrast with $\bm \Omega$, is completely determined by $\bm v_b(\bm x, t)$ and is independent of $\bm R$. This is why we use it for finding $\bm R$ in Eq.~(\ref{rotationale}). We rewrite the equation as,
\begin{eqnarray}&&\!\!\!\!\!\!\!\!\!\!\!
\frac{d\bm R}{dt}=\bm R\left(\bm W-\frac{3\tau_s}{10}\frac{d\bm W}{dt}\right), \label{ortho}
\end{eqnarray}
where we introduced the antisymmetric matrix $W$ such that,
\begin{eqnarray}&&\!\!\!\!\!\!\!\!\!\!\!
W_{lj}=-\epsilon_{lpj}\left(\int_S\!\!\frac{3\bm r\!\times\! \bm v_b(\bm r, t) dS}{8\pi a^4}\right)_p. \label{w}
\end{eqnarray}
We look for the solution of Eq.~(\ref{ortho}) to linear order in $Ro$. The zero order solution $\bm R_0$ obeys $\dot{\bm R}_0=\bm R_0\bm W$. We look for solution in the form $\bm R=(1+\delta \bm R)\bm R_0$ where $\delta \bm R\propto Ro$. We have,
\begin{eqnarray}&&\!\!\!\!\!\!\!\!\!\!\!
\frac{d\delta \bm R}{dt}=-\frac{3\tau_s}{10}\bm R_0\frac{d\bm W}{dt}\bm R_0^t=-\frac{3\tau_s}{10}\frac{d}{dt} \left[\bm R_0\bm W\bm R_0^t\right], \label{drv}
\end{eqnarray}
where we observed that,
\begin{eqnarray}&&\!\!\!\!\!\!\!\!\!\!\!
\frac{d}{dt} \left[\bm R_0\bm W\bm R_0^t\right]=\bm R_0\frac{d\bm W}{dt}\bm R_0^t,
\end{eqnarray}
which is readily verified using antisymmetry $\bm W^t=-\bm W$. We conclude integrating Eq.~(\ref{drv}) from $0$ to $t$ with $\delta\bm R(0)=0$ (implied by $R_{ij}(0)=\delta_{ij}$) that,
\begin{eqnarray}&&\!\!\!\!\!\!\!\!\!\!\!
\delta\bm R=-\frac{3\tau_s}{10}\bm R_0\left[\bm W(t)-\bm W(0)\right]\bm R_{0}^t.
\end{eqnarray}
We observe that $\delta\bm R$ is an antisymmetric matrix as it must be by the orthogonality of $\bm R$. We can write using Eq.~(\ref{w}),
\begin{eqnarray}&&\!\!\!\!\!\!\!\!\!\!\!
\delta R_{ik}=\frac{3\tau_s}{10}(R_0)_{il}(R_0)_{kj}\epsilon_{lpj}
\nonumber\\&&\!\!\!\!\!\!\!\!\!\!\!
\left(\int_S\!\!\frac{3\bm r\!\times\! \left[\bm v_b(\bm r, t)-\bm v_b(\bm r, t=0)\right] dS}{8\pi a^4}\right)_p,
\end{eqnarray}
which gives using $(R_0)_{rp}\epsilon_{rki}=(R_0)_{il}(R_0)_{kj}\epsilon_{lpj}$ that,
\begin{eqnarray}&&\!\!\!\!\!\!\!\!\!\!\!
\delta R_{ik}=\frac{3\tau_s}{10}\epsilon_{ikr} \left(\Omega^{ss}_r(t)-\Omega^{ss}_r(0)\right).\label{delta}
\end{eqnarray}
We introduced the Stone-Samuel angular velocity of inertialess swimmer in inertialess fluid,
\begin{eqnarray}&&\!\!\!\!\!\!\!\!\!\!\!\!\!\!\!\!\!
\bm \Omega^{ss}(t)=-\bm R_0\int_S\!\!\frac{3\bm r\!\times\! \bm v_b(\bm r, t)dS}{8\pi a^4}.
\end{eqnarray}
We find the swimmer's velocity using this rotation matrix from Eqs.~(\ref{swimming}) and (\ref{smro}),
\begin{eqnarray}&&\!\!\!\!\!\!\!\!\!\!\!\!\!\!\!\!\!
\bm v=\bm v^{ss}-\kappa\tau_d\frac{d\bm v^{ss}}{dt}+\frac{3\tau_s}{10} \bm v^{ss}\times\left(\bm \Omega^{ss}(t)-\bm \Omega^{ss}(0)\right),\label{swm}
\end{eqnarray}
where we used Eq.~(\ref{delta}) and introduced the Stone-Samuel velocity of inertialess swimmer in inertialess fluid \cite{stonesamuel},
\begin{eqnarray}&&\!\!\!\!\!\!\!\!\!\!\!\!\!\!\!\!\!
\bm v^{ss}=-\bm R_0\int_S \bm v_b(\bm x, t)\frac{dS}{4\pi a^2}.
\end{eqnarray}
The velocities $\bm v^{ss}$ and $\bm \Omega^{ss}$ are $\bm v_0^s$ and
$\bm \Omega^s$ in Eqs.~(\ref{swimming}) and (\ref{rt}) with rotational matrix $\bm R$ obtained neglecting the inertia. The appearance of $\bm \Omega^{ss}(0)$ is not because of infinite memory but rather because the rotation
matrix that transforms velocities from the body-fixed to the fluid frames has a reference orientation at $t=0$. Thus, the leading order correction to the swimming velocity in inertia of both the fluid and the swimmer
is derived simply from the zero order velocities, obtained neglecting the inertia.
For time-reversible stroke the scallop theorem \cite{purcell} guarantees that $\left\langle \bm v^{ss}\right\rangle=0$ so that the average velocity $\left\langle \bm v\right\rangle_{tr}$ of swimmer with time-reversible stroke obeys,
\begin{eqnarray}&&\!\!\!\!\!\!\!\!\!\!\!\!\!\!\!\!\!
\left\langle \bm v\right\rangle_{tr}=\frac{3\tau_s}{10} \left\langle  \bm v^{ss}\times\bm \Omega^{ss}\right\rangle+o(Ro). \label{tr}
\end{eqnarray}
The RHS is non-zero for asymmetric stroke demonstrating the breakdown of the scallop theorem due to inertia. We stress that this is the leading order result in $Ro$ that does not imply that time-reversible strokes for
which $\bm v^{ss}$ and $\bm \Omega^{ss}$ are parallel cannot swim by inertia. In fact, these strokes can be relevant for Volvox and non-zero swimming velocity can appear in higher-order terms.

We observe that Eqs.~(\ref{swm}), (\ref{tr}) hold also for heavy swimmers with $\sigma \tau_s \ll 1$. Thus we conclude that in this case the scallop theorem breaks down. In contrast,
\cite{gonzlauga} claimed that the scallop theorem holds also for heavy swimmers with time-reversible tangential stroke. This is because they did not consider the rotational-translational coupling for this system.

\section{Swimming at arbitrary inertia under a time-reversible stroke}

\begin{figure}
 \includegraphics[width=4.2cm]{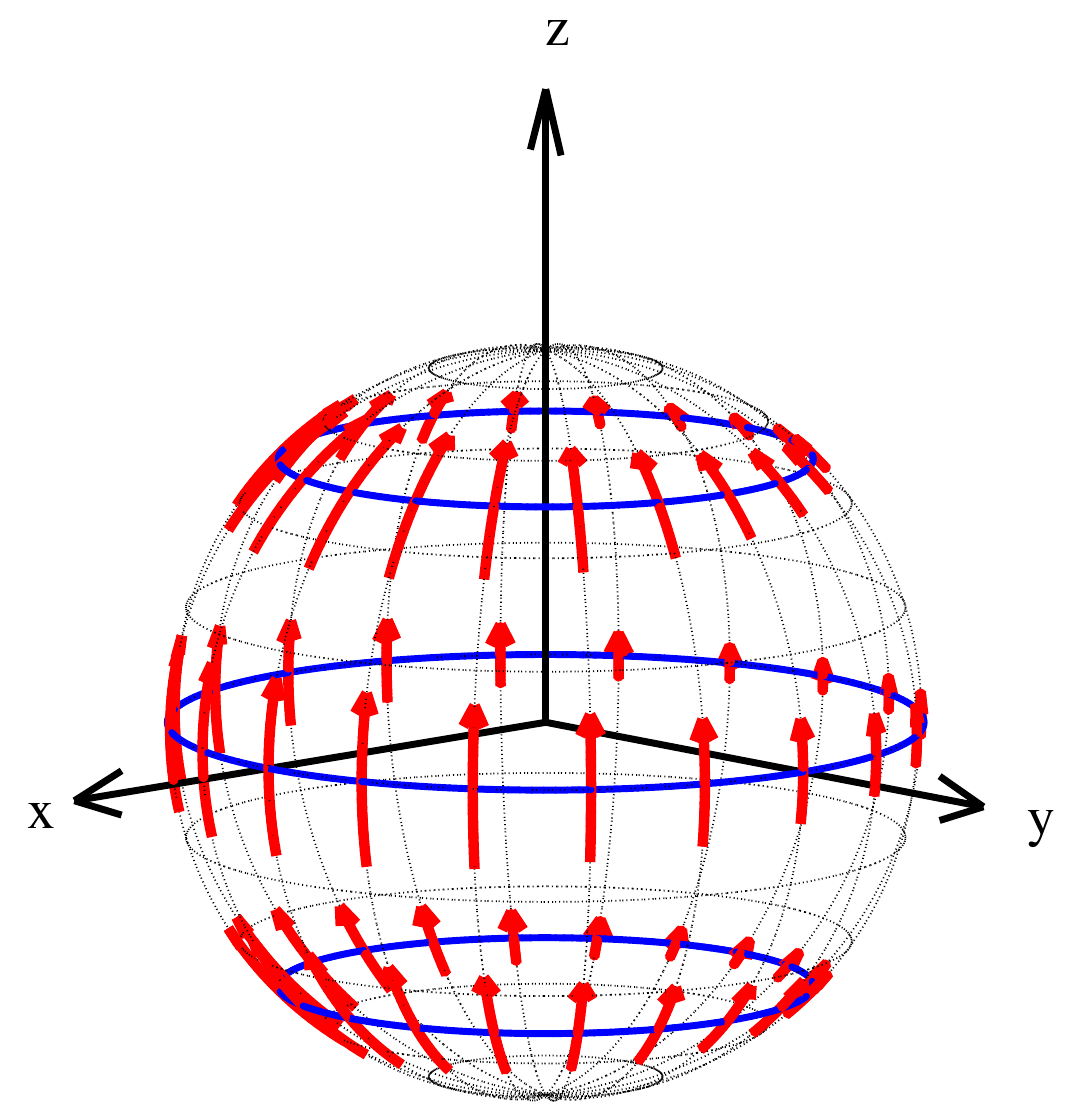}

\caption{Illustration  of the asymmetric stroke in Eq.~(\ref{str}) on a sphere. The arrows denote oscillation amplitudes of the tangential velocity distribution which are directed along longitudes and vary with $\phi$ along
latitudes. For $c=0$,
the swimmer only oscillates in pure translation along the symmetry axis $z$. For $c \neq 0$ there is an additional oscillatory rotation about the $y$ axis due to asymmetry, and the inertial rotation-translation coupling generates
nonzero net translation along the $x$ axis.
}
\label{fig:il}
\end{figure}

The observations made in the previous Section persist to higher $Ro$. We illustrate this by considering swimming due to the general asymmetric stroke given by,
\begin{eqnarray}&&\!\!\!\!\!\!\!\!
\theta(t, \theta_0, \phi_0)=\theta_0+\epsilon y(\sigma t)+c\epsilon h(\phi_0) y(\sigma t), \label{strs}
\end{eqnarray}
where $h(\phi_0)$ and $y(\sigma t)$ are some functions, $c$ is a constant and $\epsilon$ represents small amplitude of this type of cilia motion.
Below results of general calculations will be evaluated numerically for $h(\phi_0)=\cos(\phi_0)$ and $y(\sigma t)=\cos(\sigma t)$ where the stroke becomes,
\begin{eqnarray}&&\!\!\!\!\!\!\!\!
\theta(t, \theta_0, \phi_0)=\theta_0+\epsilon \cos(\sigma t)+c\epsilon \cos(\phi_0)  \cos(\sigma t). \label{str}
\end{eqnarray}
Importantly, the last term induces asymmetry that causes oscillatory rotation about $\bm {\hat y}$ direction, which, in turn, gives nonzero net
displacement. Qualitatively, equal strength time-reversible rowers are located along each longitude and the strength of the rowers depends on the azimuthal angle $\phi$, see Fig. \ref{fig:il}.

We demonstrate below that the results obtained for any asymmetry function $h(\phi_0)$ do not differ from those for $h(\phi_0)=\cos \phi_0$. In contrast, the use of different $y(t)$ can give quantitatively (but not qualitatively) different answers. We consider non-gravitational component of velocity and include gravity later.

The asymmetry function $h(\phi_0)$ describes how different the motion of cilia at different azimuthal angles is. It is a periodic function with period $2\pi$ and zero average whose Fourier series is,
\begin{eqnarray}&&\!\!\!\!\!\!\!\!
h(\phi_0)=\sum_{n=1}^{\infty}(a_n\cos(n\phi_0)+b_n\sin(n\phi_0)).\label{h}
\end{eqnarray}
We define constant $c$ in Eq.~(\ref{strs}) so that $a_1^2+b_1^2=1$. We assume that the modulation is of order one so $c$ is also of order one. The function $y$ characterizes periodic swimming stroke and has the Fourier series representation,
\begin{eqnarray}&&\!\!\!\!\!\!\!\!
y(\sigma t)=\sum_{n=1}^{\infty}(c_n\cos(n\sigma t)+d_n\sin(n\sigma t)),\label{f}
\end{eqnarray}
so that $\sigma$ is the swimming stroke frequency.
For motion given by Eq.~(\ref{strs}) the swimming stroke $\bm v_b$ has only $\theta$ component given by $v_{b\theta}(\theta, \phi)=a\epsilon\sigma \left(1 + h(\phi)c\right) y'(\sigma t)$. Thus,
\begin{eqnarray}&&
v_{bz}\!=\!-a\epsilon\sigma \left(1\!+\! h(\phi)c\right) y'(\sigma t)\sin\theta,\nonumber\\&& v_{bx}\!=\!a\epsilon\sigma \left(1\! + \!h(\phi)c\right) y'(\sigma t) \cos\phi \cos\theta,\nonumber\\&&
v_{by}\!=\!a\epsilon\sigma \left(1\! +\! h(\phi)c\right) y'(\sigma t) \sin\phi\cos\theta.
\end{eqnarray}	
It is then found,
\begin{eqnarray}&&
\bm v_{s}\!\equiv\!-\!\int_S \!\!\bm v_b(\bm x, t)\frac{dS}{4\pi a^2}\!=\!\frac{\pi}{4}a\epsilon\sigma y'(\sigma t)\hat {\bm z};\ \ \hat {\bm p}\!=\!a_1 \hat {\bm y}\!-\!b_1\hat {\bm x}, \nonumber\\&&
\bm \Omega_s\!\equiv\!-\!\int_S\!\!\frac{3\bm r\!\times\! \bm v_b(\bm r, t) dS}{8\pi a^4}\!=\!-\frac{3c\epsilon\sigma y'(\sigma t)}{4}\hat {\bm p},\label{gq}
\end{eqnarray}	
where $a_1$, $b_1$ are defined in Eq.~(\ref{h}). We observe that $\hat {\bm p}$ is a unit vector aligned with the constant rotation axis in $x-y$ plane. Using reference frame rotated in the plane so that the $ \hat {\bm p}$ has only $y-$component we find,
\begin{eqnarray}&&\!\!\!\!\!\!\!\!\!\!\!\!
\bm v_{s}=\frac{\pi}{4}a\epsilon\sigma y'(\sigma t)\hat {\bm z},\ \
\bm \Omega_s=-\frac{3c\epsilon\sigma y'(\sigma t)}{4}\hat {\bm y},
\end{eqnarray}	
which is identical to what we would obtain using $h(\phi_0)=\cos(\phi_0)$ in Eq.~(\ref{strs}).

\begin{figure}[t!]
 \begin{center}
 \includegraphics[width=.9\columnwidth]{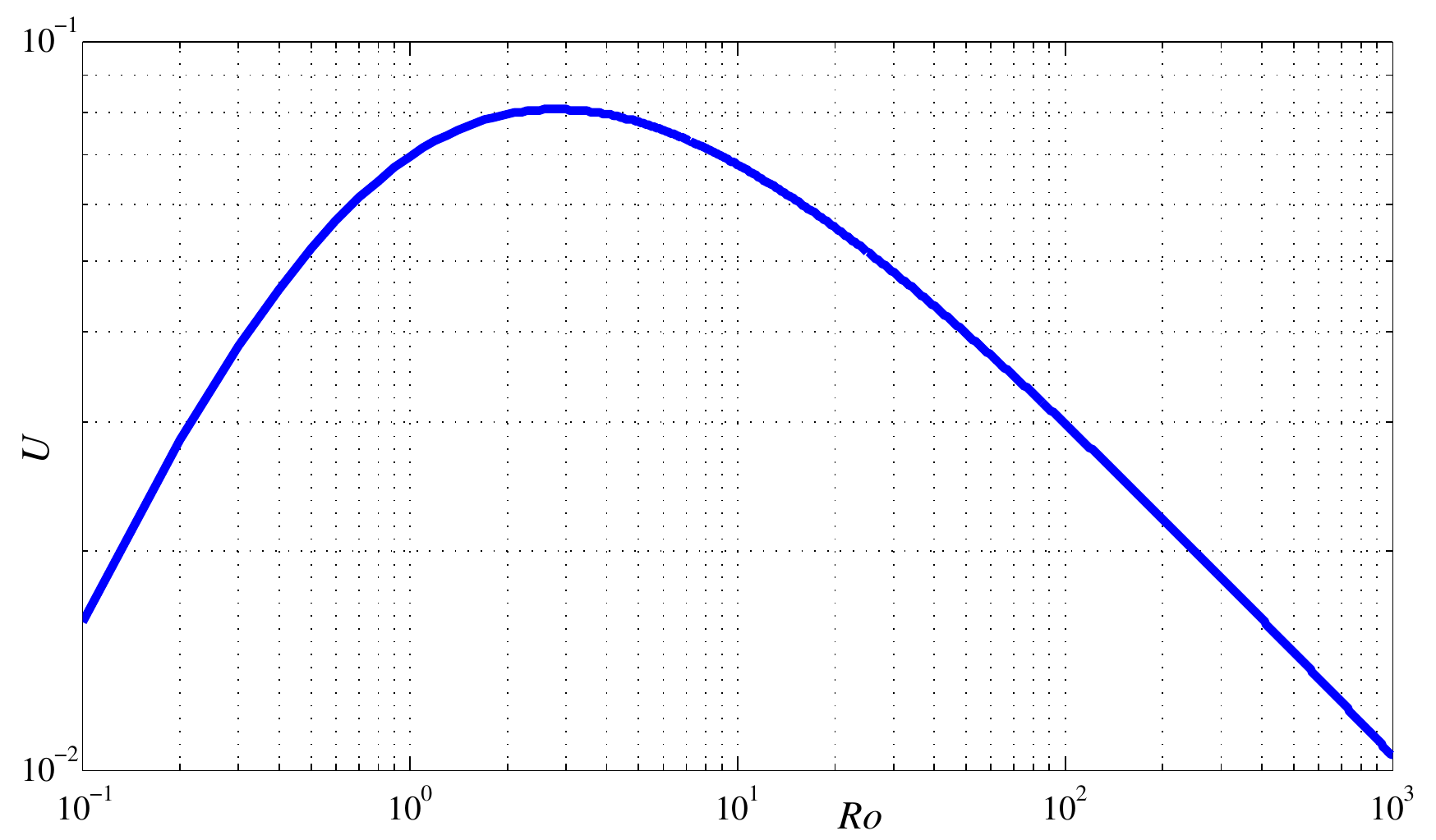}
 \end{center}
 \caption{Log-log plot of nondimensional speed $U$ as a function of Roshko number $Ro$. The maximum is reached at $Ro\approx 2.8$.}
 \label{fig}
\end{figure}

We start from considering the case of $Ro\ll 1$ where Eq.~(\ref{tr}) holds. We observe that,
\begin{eqnarray}&&\!\!\!\!\!\!\!\!\!\!\!\!\!\!\!\!\!
\bm v^{ss}\!\times\!\bm \Omega^{ss}\!=\!\bm R_0 \left(\bm v_{s}\!\times\!\bm \Omega_{s}\right)\!=\!\frac{3\pi a c\epsilon^2\sigma^2 \left(y'(\sigma t)\right)^2}{16} \bm R_0 \hat {\bm x}.
\end{eqnarray}
In the leading order in $\epsilon$ the matrix $\bm R_0$ is unit matrix since $\Omega_s\propto \epsilon$. We find from Eq.~(\ref{tr}),
\begin{eqnarray}&&\!\!\!\!\!\!\!\!\!\!\!\!\!\!\!\!\!
\left\langle \bm v\right\rangle_{tr}=\frac{9\pi a c\epsilon^2\sigma^2 \gamma\tau_d }{80} \left\langle  \left(y'(\sigma t)\right)^2\right\rangle \hat {\bm x}.
\end{eqnarray}
This gives for the stroke given by Eq.~(\ref{str}) with $y(\sigma t)=\cos(\sigma t)$ that,
\begin{eqnarray}&&\!\!\!\!\!\!\!\!\!\!\!\!\!\!\!\!\!
\left\langle \bm v\right\rangle_{tr}=\frac{\pi a c\epsilon^2\sigma \gamma\alpha^2}{80} \hat {\bm x}, \label{roshk}
\end{eqnarray}
where we defined $\alpha^2=9 \sigma\tau_d/2$. Returning to the case of arbitrary $Ro$, we have from Eq.~(\ref{rp}) that,
\begin{eqnarray}&&\!\!\!\!\!\!\!\!\!\!\!
\bm {\hat \Omega}(\omega)=\frac{\left(\bm R\bm \Omega_s\right)(\omega)}{1-iJ\omega T_0^{-1}(\omega)}=-\frac{3c\epsilon\sigma}{4}\frac{\left(y'(\sigma t)\bm R\hat {\bm y}\right)(\omega)}{1-iJ\omega T_0^{-1}(\omega)},
\end{eqnarray}
where in this formula we designate Fourier transform of some function $q(t)$ by $(q)(\omega)$. We observe that since the rotation is around $y-$axis then $\bm R\hat {\bm y}=\hat {\bm y}$. We find,
\begin{eqnarray}&&\!\!\!\!\!\!\!\!
\bm {\hat \Omega}(\omega)=\frac{3i\pi c\epsilon\omega\hat {\bm y}}{4}\sum_{n=1}^{\infty}\left(\frac{ c_n\left(\delta(\omega+n\sigma)+\delta(\omega-n\sigma)\right)}{1-iJ\omega T_0^{-1}(\omega)}
\right.\nonumber\\&&\!\!\!\!\!\!\!\!\left.
-\frac{i d_n\left(\delta(\omega+n\sigma)-\delta(\omega-n\sigma)\right)}{1-iJ\omega T_0^{-1}(\omega)}
\right),
\end{eqnarray}
where we used Eq.~(\ref{f}). Inverse Fourier transform gives,
\begin{eqnarray}&&\!\!\!\!\!\!\!\!
\bm  \Omega(t)\!=\!\frac{3 c\epsilon  \sigma \hat {\bm y}}{4}\sum_{n=1}^{\infty}\left(n c_n {\cal I}\left(\frac{\exp(in\sigma t)}{1\!+\!iJn\sigma T_0^{-1}(-n\sigma)}\right)
\right.\nonumber\\&&\!\!\!\!\!\!\!\!\left.
-nd_n {\cal R}\left(\frac{\exp(in\sigma t)}{1\!+\!iJn\sigma T_0^{-1}(-n\sigma)}\right)\right),
\end{eqnarray}
where we used $T_0(-\omega)=T_0^{*}(-\omega)$ and ${\cal I}$ stands for imaginary part. Introducing rotation angle $\psi(t)=\int_0^t \Omega(t')dt'$ we find,
\begin{eqnarray}&&\!\!\!\!\!\!\!\!
\psi(t)\!=\!\frac{3c\epsilon}{4} \sum_{n=1}^{\infty}\left(c_n {\cal R}\left(\frac{1-\exp(in\sigma t)}{1\!+\!iJn\sigma T_0^{-1}(-n\sigma)}\right)
\right.\nonumber\\&&\!\!\!\!\!\!\!\!\left.
+d_n {\cal I}\left(\frac{1-\exp(in\sigma t)}{1\!+\!iJn\sigma T_0^{-1}(-n\sigma)}\right)\right).
\end{eqnarray}
We have that up to quadratic order in $\epsilon$,
\begin{eqnarray}&&\!\!\!\!\!\!\!\!\!\!\!
\bm R(t)\bm v_s(t)=\frac{\pi }{4}a\epsilon\sigma y'(\sigma t)\psi(t) \hat {\bm x}+O\left(\epsilon^3\right).\label{R}
\end{eqnarray}
We find for the time average velocity using that it equals the time average of $\bm R(t)\bm v_s(t)$ giving $\langle \bm v\rangle=\pi a\epsilon\sigma \langle y'(\sigma t)\psi(t)\rangle \hat {\bm x}/4$ that,
\begin{eqnarray}&&\!\!\!\!\!\!\!\!\!\!\!
\langle \bm v\rangle
=-\frac{3\pi ca\epsilon^2\sigma \hat {\bm x}}{16}
\sum_{n=1}^{\infty}\left(c_n {\cal R}\left\langle\frac{\exp(in\sigma t)y'(\sigma t)}{1\!+\!iJn\sigma T_0^{-1}(-n\sigma)}\right\rangle\right.\nonumber\\&&\!\!\!\!\!\!\!\!\left.+
d_n {\cal I}\left\langle \frac{\exp(in\sigma t)y'(\sigma t)}{1\!+\!iJn\sigma T_0^{-1}(-n\sigma)}\right\rangle\right). \nonumber
\end{eqnarray}
We find performing time averaging,
\begin{eqnarray}&&\!\!\!\!\!\!\!\!\!\!\!
\langle \bm v\rangle=\frac{3\pi ca\epsilon^2\sigma  \hat {\bm x}}{32}\sum_{n=1}^{\infty} {\cal R}\left(\frac{i (c_n^2+d_n^2) }{1\!+\!iJn\sigma T_0^{-1}(-n\sigma)}\right).
\end{eqnarray}
We observe that,
\begin{eqnarray}&&\!\!\!\!\!\!\!\!\!\!\!
{\cal R}\left(\frac{i }{1\!+\!iJn\sigma T_0^{-1}(-n\sigma)}\right)={\cal R}\left[\left(\frac{T_0(-n\sigma)}{Jn\sigma}\!+\!i\right)^{-1}\right].\nonumber
\end{eqnarray}
We have from Eq.~(\ref{t0}) that,
\begin{eqnarray}&&\!\!\!\!\!\!\!\!\!\!\!\!\!\!\!\!\!
\frac{T_0(-n\sigma)}{Jn\sigma}=\frac{15}{2\gamma \alpha_n^2} +\frac{5 i }{\gamma\left(1\!+\!\alpha_n\!+\!i\alpha_n\right)}
,
\end{eqnarray}
where we introduced $\alpha_n=3\sqrt{n Ro/2}$. We find,
\begin{eqnarray}&&
\langle \bm v\rangle=U_0a\sigma\hat {\bm x}+2(\gamma-1)\bm g\tau_d;\ \ \
U_0=\frac{3\pi c\epsilon^2  }{32}\\&&\cdot
\sum_{n=1}^{\infty} (c_n^2+d_n^2)
{\cal R}\left[\left(\frac{15}{2\gamma \alpha_n^2}\!+\!i\!+\!\frac{5i}{\gamma\left(1\!+\!\alpha_n\!+\!i\alpha_n\right)}\right)^{-1}\right].\nonumber
\end{eqnarray}
where we introduced the velocity $U_0$ that determines ``displacement in body sizes per stroke'' and included the sedimentation velocity due to gravity.
The corresponding formula for the case of Eq.~(\ref{str}) is obtained by setting all $c_n$ and $d_n$ to zero except for $c_1=1$ which gives,
\begin{eqnarray}&&\!\!\!\!\!\!\!\!\!\!\!
U\!=\!\frac{3\pi c\epsilon^2}{32}{\cal R}\!\left[\left(\frac{15}{2\gamma \alpha^2}\!+\!i\!+\!\frac{5i}{\gamma\left(1\!+\!\alpha\!+\!i\alpha\right)}\right)^{-1}\right]\!.\label{cos}
\end{eqnarray}
where $\alpha\!=\!3\sqrt{Ro/2}$ was defined after Eq.~(\ref{roshk}). This formula reproduces Eq.~(\ref{roshk}) in the small Roshko number limit.
We study $U$ as a function of the free parameters of the swimming stroke $\gamma$, $\sigma$ and $a$. Writing Eq.~(\ref{cos}) explicitly as a real-valued function gives,
\begin{eqnarray}&&\!\!\!\!\!\!
U\!= \!\left|\frac{15}{2\gamma \alpha^2}\!+\!\frac{5\alpha}{\gamma\left(1\!+\!2\alpha\!+\!2\alpha^2\right)}\!+\!i\!+\!\frac{5i\left(1\!+\!\alpha\right)}{\gamma\left(1\!+\!2\alpha\!+\!2\alpha^2\right)}\right|^{-2}
\nonumber\\&&\!\!\!\!\!\!
\cdot \frac{3\pi c\epsilon^2}{32}\left(\frac{15}{2\gamma \alpha^2}+ \frac{5\alpha}{\gamma\left(1\!+2\alpha+2\alpha^2\right)}\right).\nonumber
\end{eqnarray}
We find that dimensionless velocity factorizes as,
\begin{eqnarray}&&\!\!\!\!\!\!\!\!\!\!\!
U=\frac{3\pi c \epsilon^2}{32}{\tilde U}(\gamma, \alpha),
\end{eqnarray}
where the dimensionless function ${\tilde U}$ of two dimensionless numbers $\gamma$ and $\alpha$ is,
\begin{eqnarray}&&\!\!\!\!\!\!\!\!\!\!\!\!\!\!\!\!
{\tilde U}\!=\!\gamma\left(1\!+\!2\alpha\!+\!2\alpha^2\right) \left(15\left(1\!+\!1/\alpha\!+\!1/(2\alpha^2)\right)\!+\!5\alpha\right)/D,\label{fic}
\end{eqnarray}
with the denominator $D$,
\begin{eqnarray}&&\!\!\!\!\!\!
D=\left(15\left(1\!+1/\alpha+1/(2\alpha^2)\right)+5\alpha\right)^2\nonumber\\&&\!\!\!\!\!\!
+\left(\gamma\left(1\!+2\alpha+2\alpha^2\right)+5\left(1\!+\!\alpha\right)\right)^2.
\end{eqnarray}
Fig. \ref{fig}
shows a log-log plot of $U$ as a function of Roshko number $Ro$ for $\gamma=1$ where sedimentation velocity vanishes.
Interestingly, the speed $U$ vanishes at the limits $Ro \to \{0,\infty\}$ and attains a maximum at an intermediate
value of $Ro \approx 2.8$ (vanishing at $Ro\to 0$ is due to the scallop theorem \cite{purcell} and at $Ro\to\infty$ due to large mass). This can be interpreted either as an optimal flapping frequency $\sigma$, or as an optimal body radius $a$ for a fixed frequency. Remarkably, setting physical values of $\sigma=200\ rad/sec$  and $\nu =10^{-6}  m^2/sec$ for Volvox in water gives an optimal radius of $a\approx 355\mu m$, which falls well within the size range of large Volvox colonies \cite{Goldstein}.

We also observe that the dependence on the density of the swimmer has the form $U(\gamma)=\gamma/(b_2\gamma^2+b_1\gamma+b_0)$  where $b_i$ are functions of $Ro$ only. The corresponding definitions of $b_i$ can be readily obtained from Eq.~(\ref{fic}). Elementary calculus of $U'(\gamma)=0$ gives that $U$ considered as a function of density of the swimmer has a maximum at an optimal density ratio of,
\begin{eqnarray}&&\!\!\!\!\!\!\!\!\!\!\!
\gamma^*\!=\!\frac{5\sqrt{\left(3/2\alpha^2\!+\!3/\alpha\!+\!3\!+\!\alpha\right)^2\!+\!\left(1\!+\!\alpha\right)^2}}{1\!+\!2\alpha\!+\!2\alpha^2}.
\end{eqnarray}
While for physical values of Volvox the optimal value is $\gamma \sim 10$, this value results in a large sedimentation velocity in Eq.~(\ref{d}) which can hardly be overcome by
the strokes. Therefore, Volvox typically tend to a density ratio of nearly neutral buoyancy ($\gamma =1.003$, cf. \cite{Goldstein}) for which sedimentation is nonzero yet small and
the above optimum has less implications.

We consider Eq.~(\ref{cos}) at the size ranges of $100$ - $500$ microns that is typical for Volvox and at fixed $\sigma$ (the period of flagella motion depends weakly on the size). We find that
dimensionless swimming velocity depends on the dimensionsless parameters of the stroke as $\epsilon^2$ times order one constant $c$ times a numerical factor. This type of parameteric dependence is quite universal.
For instance, for the stroke $\theta=\theta_0+\epsilon\cos(n\theta_0-\omega t)$ considered in \cite{stonesamuel} we have $U\propto n\epsilon^2$ . The same stroke combined with small radial deformations is considered to model Volvox,
see e. g. \cite{swirl}. The radial deformations do not bring strong change in the parameteric dependence: we have $U\propto \epsilon^2$ with proportionality coefficient depending on $n$ and the ratio of amplitudes of angular and
radial motions. We conclude that the flapping stroke prescribed by Eq.~(\ref{strs}) generates swimming velocity that is quite similar to that for irreversible strokes. This is of course because inertia is of order one, $Ro\sim 1$
so that the scallop theorem's breakdown is of order one. Since the radial displacements would probably not destroy the phenomenon then this way of self-propulsion could be used by Volvox for reaching velocities similar
to those achieved by irreversible strokes. This way of swimming, if realized, could be done during turns toward/away from external stimuli such as light in phototaxis \cite{photo,paklauga}. It might also occur
in emergency cell swimming similar to escape of Paramecium from heat source where time-reversible stroke was observed \cite{para,ther}. Whether this type of swimming is really used by Volvox or other microswimmer for the actual
motion is to be decided by future observations from experimental measurements.

\section{Summary and discussion}

We have studied the motion of inertial squirmers under unsteady Stokes equation. Using reciprocal theorem, we derived the swimmer's translation and rotation under any given stroke of tangential deformation.
Our analysis extends the well-known work \cite{stonesamuel} of Stone and Samuel from inertialess to inertial case. It also generalizes previous works on inertial squirmers \cite{ardekani} which considered only axisymmetric strokes
where rotation is either zero or decoupled from translation \cite{swirl}. As a consequence, we show that an asymmetric time-reversible stroke can lead to net propulsion through dynamic coupling between rotation and translation.
For small inertia this coupling is described by the vector product of inertialess translational and rotational velocities. If the product has non-zero time-average then the net propulsion velocity is finite even for time-reversible
stroke. For a model of asymmetric stroke, the normalized swimming speed is maximized for intermediate values of $O(1)$ Roshko number which falls well within realistic range of large Volvox colonies. For $Ro \to 0$, our results
reduce to those of inertialess swimmer in \cite{stonesamuel}. For $Ro \to \infty$, $U$ decays as $1/\sqrt{Ro}$. We conjecture that swimming optimization at $Ro\sim 1$ can be one of the reasons for the cutoff in size of Volvox
colonies at $a \approx 500 \mu m$ \cite{Goldstein}.

We now briefly discuss some limitations of our work and suggest possible directions for future extension. Our current work is limited to tangential strokes, whereas Volvox's strokes involve also radial component, which must be included in future extensions of the analysis for realistic description. For axisymmetric strokes,
time-reversible deformation with nonzero radial component can lead to net propulsion of inertial squirmers \cite{gonzlauga,ishimoto}. Nevertheless, inclusion of the radial stroke seemingly would not destroy the asymmetric
swimming mechanism displayed here. Indeed, considering surface deformations as a sum of radial and tangential displacements we would find the sum of the corresponding contributions also in the swimming velocity. The velocity
would include also the terms due to the coupling of radial and tangential deformations. There seems to be no reason why the terms involving radial displacements would cancel the terms due to tangential displacements, destroying
the effect described here. Of course only the complete calculation can fully prove this point, which is left as an open challenge.
We plan to compare the two mechanisms of swimming (our mechanism and axisymmetric time-reversible deformation with nonzero radial component) using a full solution of the flow in the vicinity of the deforming boundary for
general strokes. The general solution will also enable calculation of mechanical energy dissipation, which will give another criterion for stroke optimization.

We did not consider here the effects of bottom-heaviness (center-of-mass displacement from the sphere's center). This could produce coupling between rotation, caused by the gravitational torque, and inertial translation similar to
that considered in this work. The study of this coupling is of interest and left for future work.
Finally, a longer term goal is to use these improvements to the theoretical model in order to match it with experimental measurements of swimming Volvox. This will help in quantifying the true contribution of inertial effects to Volvox motion.

Large organisms often use time-reversible stroke for swimming while small ones do not, since they obey the scallop theorem \cite{purcell}. When inertia becomes not completely negligible the time-reversible stroke becomes a
possible way of swimming and it can be used as in the case of Paramecium \cite{para,ther}. It seems of interest to study how the efficiency of the time-reversible stroke improves at
growing inertia of the organisms and at which time-reversible stroke becomes advantageous. Our work provides a step in this direction of study.

Our main discovery in this work is that the spherical squirmer model contains strokes in which swimming can occur by inertial coupling of rotation and translation. In this context there rises a question: is the spherical
squirmer model, possibly with inclusion of most general small deformations of the shape and the effects
of inertia, able to shed light on the reasons for the rotation of Volvox in
nature? Currently the only attempt at incorporation of the Volvox rotation in the theory is done with the help of axisymmetric (independent of the azimuthal angle)
swirl \cite{swirl}. In this frame rotation and translation are uncoupled, leaving the reasons for rotation unclear. The authors of \cite{swirl} proposed that the modeling of Volvox interaction with the fluid by a continuous boundary whose shape is fully controlled by the swimmer, can be invalid. In fact, our work demonstrates that
not all the phenomena that can be captured by the model are exhausted. The possibility of coupling rotation to translation can give new reasons for rotating (as it did in the study of phototaxis \cite{photo,paklauga}).
Unfortunately, the stroke introduced in this work is currently detached from the data, and more fully time-resolved measurements of strokes of real Volvox are needed for progress. The challenge of a proper theoretical understanding of Volvox rotation
seems to stand open and requires further work.

\section*{Acknowledgments}

This work has been supported by the Israel Science Foundation (ISF), under grant no. $567/14$.

{}

\end{document}